\documentclass[aps,prl,twocolumn]{revtex4}
\usepackage{graphicx}
\begin{document}

\title{Quantum critical 5f-electrons avoid singularities in U(Ru,Rh)$_2$Si$_2$}
\author{A. V. Silhanek,$^1$ N. Harrison,$^1$\footnote{Correspondence and requests for materials should be addressed to N.H. (e-mail: nharrison@lanl.gov)} C. D. Batista,$^2$ M. Jaime,$^1$ A. Lacerda,$^1$ H. Amitsuka,$^3$ and J. A. Mydosh$^{4,5}$}
\address{$^1$National High Magnetic Field Laboratory, Los Alamos National Laboratory, MS E536,
Los Alamos, NM 87545, USA\\
$^2$Los Alamos National Laboratory, MS B262, Los Alamos, NM 87545, USA\\
$^3$Graduate School of Science, Hokkaido University, N10W8 Sapporo 060-0810,Japan\\
$^{4}$Kamerlingh Onnes Laboratory, Leiden University, 2300RA Leiden, TheNetherlands\\
$^{5}$Max-Planck-Institut f\"{u}r Chemische Physik fester Stoffe, 01187 Dresden, Germany}

\date{\today}

\begin{abstract}

We present specific heat measurements of 4\% Rh-doped URu$_2$Si$_2$ at magnetic fields above the proposed metamagnetic  transition field $H_{\rm m}\gtrsim$~34~T, revealing striking similarities to the isotructural Ce analog CeRu$_2$Si$_2$, suggesting that strongly renormalized hybridized band models apply equally well to both systems. The vanishing bandwidths as $H\rightarrow H_{\rm m}$ are consistent with a putative quantum critical point close to $H_{\rm m}$. The existence of a phase transition into an ordered phase in the vicinity of $H_{\rm m}$ for 4\% Rh-doped URu$_2$Si$_2$, but not for CeRu$_2$Si$_2$, is consistent with a stronger super-exchange in the case of the U 5$f$ system, with irreversible processes at the transition revealing a strong coupling of the $5f$ orbitals to the lattice, most suggestive of orbital or electric quadrupolar order.

\end{abstract}

\pacs{PACS numbers: 71.27.+a, 75.30.Kz, 75.45.+j}

\maketitle
A quantum-critical point is a singular feature in the phase diagram of matter at the absolute zero of temperature~\cite{sachdev-1,sachdev-2}. At this point, the quantum fluctuations that result from the Heisenberg uncertainty principle acquire a divergent characteristic length~\cite{sachdev-1,sachdev-2}. Quantum fluctuations originating from this singularity influence the physical properties of matter over an expanding region of phase space (pressure, magnetic field and chemical doping) as the temperature increases~\cite{coleman-1}. Several unexpected ordered states in strongly correlated matter, including unconventional superconductivity in $f$-electron intermetallics~\cite{coleman-2} and $d$-electron oxides~\cite{alff,chakravarty}, occur in the vicinity of a magnetic quantum critical point. Consequently, theoretical models have focused on the role of symmetry-breaking quantum-critical points in their formation~\cite{coleman-1,coleman-2,alff}. 

In this paper, we present the first direct thermodynamic evidence for the avoidance of a non-symmetry-breaking quantum-critical point by the creation of a new low temperature ordered state. In this case, quantum criticality is caused by metamagnetism induced by strong magnetic fields in 4\% Rh-doped URu$_2$Si$_2$~\cite{kim}, where Rh substitutes Ru so as to yield URu$_{1.92}$Rh$_{0.08}$Si$_2$. Our specific heat measurements reveal the presence of narrow 5$f$-bands at high magnetic fields, whose entropy then drops abruptly upon ordering at a distinct first-order phase transition. Irreversibility of the transition yields that it is of first order, suggestive of a strong coupling of the ordering $5f$-electron degrees of freedom to the lattice.

URu$_2$Si$_2$~\cite{palstra} and its Rh-doped alloys~\cite{miyako} belong to a class of strongly-correlated metals~\cite{fazekas}, that includes CeRu$_2$Si$_2$~\cite{flouquet}, UPt$_3$~\cite{frings} and Sr$_3$Ru$_2$O$_7$~\cite{perry}, in which the $d$- or $f$-electrons are itinerant (i.e. they contribute to the metallic properties of the material) but are on the threshold of becoming localized and giving rise to magnetism. By coupling directly to their spin degrees of freedom, strong magnetic fields can coax the $d$- or $f$-electrons into a polarized state. `Metamagnetism' results when this transformation occurs abruptly at a critical magnetic field $H_{\rm m}$, as depicted in Fig. 1a. Should $H_{\rm m}$ evolve from a crossover at finite temperatures into a phase transition (analogous to that of a liquid-gas phase transition) very close to absolute zero~\cite{grigera}, it then develops all of the characteristics of an isolated non-symmetry-breaking quantum critical point~\cite{millis}, as depicted in Fig. 1b. Stoichiometric URu$_2$Si$_2$, CeRu$_2$Si$_2$ and Sr$_3$Ru$_2$O$_7$~\cite{perry} are sufficiently close to quantum criticality at $H_{\rm m}$ for their physical properties to be strongly influenced by fluctuations at temperatures $T\gtrsim$~1~K.

Being composed of 5$f$-electrons that have properties intermediate between those of the $d$-electrons in transition metal oxides and 4$f$-electrons in rare earth intermetallics~\cite{smith}, actinide intermetallics such as U(Ru,Rh)$_2$Si$_2$ occupy a unique vantage point for understanding the dynamics of quantum criticality in the formation of new states. While not as spatially extended as $d$-orbitals, the 5$f$-orbitals of U exhibit a sizeable degree of super-exchange between neighbouring U sites~\cite{fazekas}, greatly increasing the likelihood of an ordered state over its 4$f$ analogue, Ce. As with 4$f$-electrons, however, the on-site Coulomb repulsion between 5$f$-electrons is sufficiently strong to facilitate the formation of renormalized (narrow) bands upon their hybridization with regular conduction bands~\cite{hewson}. This has two immediate benefits: firstly, the narrow 5$f$-band can be completely polarized by magnetic fields that are available in the laboratory. $H_{\rm m}$ in URu$_{1.92}$Rh$_{0.08}$Si$_2$ occurs at 34~$<\mu_0H<$~37~T~\cite{kim}, bringing it well within the limits ($\sim 45$ T) of the highest available static magnetic fields. Secondly, the Fermi temperature ($T^*<$~20~K) of these quasi-particle bands is significantly lower than the characteristic Debye temperature ($T_\theta\gg$~30~K) of the phonons (or lattice vibrations)~\cite{amitsuka}, making the magnetic field-dependent degrees of freedom of these polarized bands readily accessible to fundamental thermodynamics probes such as the specific heat~\cite{jaime}.  By comparison, the comparatively large energy scale for $d$-bands in the cuprates~\cite{alff} continues to be a major impediment in attempts to identify a possible link between quantum criticality and phase formation in the high temperature superconductors. 

Figure 2a shows the temperature dependence of the specific heat of URu$_{1.92}$Rh$_{0.08}$Si$_2$ divided by temperature $C_p/T$ at several values of the magnetic field $H$. The relatively small contribution from the phonons for $T<$~20~K (estimated from non-magnetic ThRu$_2$Si$_2$)~\cite{amitsuka} implies that $C_p/T$ in Fig. 2a is dominated by the electronic contribution, having an appearance similar to that of a Schottky anomaly, but with an additional quadratic tail at low temperatures. In order to understand this behaviour for $C_p/T$, it is instructive to compare it with similar data obtained by van der Meulen {\it et al}~\cite{vandermeulen} for the isostructural Ce analog CeRu$_2$Si$_2$ (also a metamagnet) shown in Fig. 2b, for which super-exchange interactions between neighbouring 4$f$-sites are expected to be comparatively unimportant~\cite{fazekas}. The overall electronic structure of CeRu$_2$Si$_2$ at fields $H>H_{\rm m}$ has already been shown to be consistent with the general theoretical framework of the Anderson lattice model in which the 4$f^1$ magnetic doublets are hybridized with a broad conduction band~\cite{evans,edwards}. Following the qualitative picture of Edwards and Green~\cite{edwards} for the evolution of the quasi-particle up and down bands, we can approximate the corresponding density of electronic states (per unit of energy) by 
\begin{equation}
D(\varepsilon) \approx D_0 \left( 1 + \Sigma_{\sigma} \frac{q_{\sigma}V^2}{(\varepsilon-\mu-\varepsilon_\sigma)^2+\Delta^2} \right).
\label{eq1}
\end{equation}
$D_0$ is the density of states of the broad unperturbed conduction band, $V$ is the hybridization potential, while $\varepsilon_\sigma$ is the energy shift of each quasi-particle band due to the interplay between the Kondo interaction and the Zeeman (or magnetic field) coupling at fields $H > H_{\rm m}$.   $q_\sigma^{-1}$ represents the extent to which the density of electronic states is renormalized by the strong Coulomb interactions between $f$-electrons\cite{edwards}. The parameter  $\Delta =  \pi q_\sigma V^2/D_0$ has to be adjusted to accommodate one electron per formula unit, becoming the effective width of the hybridized bands\cite{edwards}. Although the important spectral weight around the bare $f$-level is missing in this approach, it is entirely adequate for calculating $C_v$ of CeRu$_2$Si$_2$ in the limit $|\varepsilon_\sigma-\mu|>\Delta$. Hence, Equation \ref{eq1} develops a simple Lorentzian form.

Figures 2c, 2d  illustrate the results of fits for $C_p/T$ versus $T$ (shown as solid lines in Figs. 2a and b) where $C_p\sim C_v=T\partial^2F/ \partial T^2|_v$ is calculated numerically from the free energy $F=\int_{-\infty}^{\infty}D(\varepsilon)\ln(1+\exp(\mu-\varepsilon)/k_{\rm B}T){\rm d}\varepsilon$~\cite{schotte} and where each spin component  $\sigma=\pm 1/2$ is considered independent in the present hybridized-band approximation. These fits confirm that one spin component of the hybridized band, which we shall assume to be spin-up, $\varepsilon_{-1/2}$~\cite{evans,edwards}, is closer to the chemical potential $\mu$ and crosses $\mu$ at $H\sim H_{\rm m}$, in accordance with theoretical expectations for the Anderson lattice~\cite{evans,edwards}.

The close similarity of Fig. 2a to 2b and Fig. 2c to 2d implies that there exists an extensive range of magnetic fields and temperatures for which the hybridized band model applies equally well to the 5$f$-electrons in URu$_{1.92}$Rh$_{0.08}$Si$_2$ as it does to the 4$f$-electrons in CeRu$_2$Si$_2$. It also implies that the orbital manifold of U in U(Ru,Rh)$_2$Si$_2$ is a doublet, as opposed to a singlet, which has been one of the pivotal areas of debate in attempts to understand the Hidden Order phase in pure URu$_2$Si$_2$ (suppressed in Rh-doped URu$_2$Si$_2$)~\cite{palstra,miyako,fazekas,barzykin,santini,ohkawa}. The similarity of the form of $C_p/T$ in Fig.~2 to that in pure URu$_2$Si$_2$~\cite{jaime}, implies that the existence of this doublet is not significantly affected by small amount of Rh-doping.
Figure 3 further shows that the fitted bandwidth $\Delta$  for both URu$_{1.92}$Rh$_{0.08}$Si$_2$ and CeRu$_2$Si$_2$ plotted versus $H-H_{\rm m}$ , is the same for both systems, within experimental uncertainty. For both systems, $\Delta \propto q_\sigma$, revealing that the bands become progressively more narrow as the spin fluctuations intensify,  since $q_\sigma^{-1} \propto |H-H_{\rm m}|^{-1}$ exhibits a divergent behaviour near $H_{\rm m}$. The dotted line in Fig. 3 shows an independent estimate of the Fermi temperature $T^*$ of the quasi-particle bands obtained from magnetotransport measurements on URu$_{1.92}$Rh$_{0.08}$Si$_2$~\cite{kim}. Its consistency with $\Delta$ provides the first confirmation of a direct correlation between features observed in the electrical resistivity and the hybridized bandwidth~\cite{evans,edwards}. 

While URu$_{1.92}$Rh$_{0.08}$Si$_2$ and CeRu$_2$Si$_2$ possess many similarities for $H-H_{\rm m}\gtrsim$~4~T, significant differences emerge as $H\rightarrow H_{\rm m}$, as shown in Fig. 4. This can been seen rather directly in URu$_{1.92}$Rh$_{0.08}$Si$_2$ as soon as $\mu_0H$ is reduced from 38~T to 37.5~T in Fig. 4a. At temperatures above $\sim$~ 6~K, $C_p/T$ at  $\mu_0H\sim$~37.5~T conforms to the solid curve calculated using fitting results for $\Delta$, $\varepsilon_{+1/2}$ and $\varepsilon_{-1/2}$ extrapolated from $\mu_0H\geq$~38~T in Fig. 2c. Thus, at higher temperatures, the specific heat of URu$_{1.92}$Rh$_{0.08}$Si$_2$ at 37.5~T continues to be consistent with quasi-particle bands that become progressively more narrow and closer to $\mu$ as $H \rightarrow H_{\rm m}$. At temperatures below 6~K, however, a significant redistribution of entropy occurs with respect to the calculated curve (cyan shaded area), establishing rather conclusively that the same 5$f$-electrons involved in the formation of the quasi-particle bands condense into a new state at low temperatures. The sharp anomaly at $\sim$~4.8~K at 37.5~K provides unambiguous evidence for the existence of a phase transition. Figure 4b further shows that the amount of energy required to heat the sample during the specific heat measurement differs considerably between initial (open symbols) and subsequent (filled symbols) $\approx$~0.1~K cycles of the temperature, using the relaxation method. Hence, the actual phase transformation itself is an energetically costly process, resulting in considerable hysteretic losses characteristic of a first order phase transition. This observation closely reproduces that observed at the valence transition in YbInCu$_4$~\cite{sarrao,silhanek}, at which a change in the orbital manifold of the $f$-electrons is coupled to the lattice parameters~\cite{dallera}. This finding in URu$_{1.92}$Rh$_{0.08}$Si$_2$ is most suggestive of orbital or electric quadrupolar order~\cite{ohkawa}.

CeRu$_2$Si$_2$, by contrast, does not transform into a new state at low temperatures\cite{flouquet}. The similarity in the intensity of the fluctuations in the two systems suggests that while they play a crucial role in driving the system towards instability at $H_{\rm m}$, the increased tendency for direct or super-exchange between 5$f$-orbitals compared to 4$f$-orbitals appears to be the decisive factor in whether a new ordered phase actually occurs. The very appearance of ordered phases in U(Ru,Rh)$_2$Si$_2$ in connection with an isolated non-symmetry-breaking quantum-critical point (as opposed to one that is symmetry breaking), suggests that the tendency to form new states of matter is ubiquitous to both forms of quantum criticality. Such a finding may have far reaching implications, for if it proves to be generic (i.e. not unique to metamagnetism) it would eliminate the need to associate the pseudogap regime in the cuprates with a true broken-symmetry phase~\cite{alff,balakirev}. Finally, if it is the exchange between the orbitals that ultimately optimizes conditions for the formation of an ordered phase, this would help to explain the common trend in maximum ordering transition temperatures in progressing from 4$f$ to 5$f$ to $d$-electrons. 

\acknowledgments
This work was performed under the auspices of the National Science Foundation, the Department of Energy (US) and the State of Florida.
\bibliographystyle{prsty}

\begin{figure}[hctb]
  \centering
  \caption{A schematic of metamagnetic quantum criticality. (a) shows an illustration of the inflection point in the magnetization at $H_{\rm m}$ which is anticipated to acquire an infinite slope at $T = 0$ (blue line), but which becomes broadened at finite temperatures $T > 0$ (red line). (b) shows the resultant magnetic field $H$ versus temperature $T$ phase diagram, with a heavy Fermi liquid (strongly correlated electron fluid) region at $H<H_{\rm m}$ and polarized Fermi liquid region at $H>H_{\rm m}$. In the ideal case, the system is a non-Fermi liquid at $T=$~0 at $H=H_{\rm m}$ (black spot). At finite temperatures $T>$~0, the region of phase space occupied by the non-Fermi liquid expands, giving rise to the recognizable funnel shape and, in doing so, relieves the necessity for the quantum criticality being precisely tuned by pressure, chemical doping or magnetic field at $T=$~0.}
  \label{fig1}
\end{figure}

\begin{figure}[hctb]
  \centering
  \caption{The measured specific heat divided by temperature $C_p/T$ together with the results of model fitting. (a) shows $C_p/T$ for URu$_{1.92}$Rh$_{0.08}$Si$_2$ versus temperature $T$ at several different values of the magnetic field $H > H_{\rm m}$ depicted using different symbols and colours as indicated. Solid lines indicate the fits to the hybridized band model. (b) gives similar results for CeRu$_2$Si$_2$. (c) shows fitted values for the position of the spin-up $\varepsilon_{-1/2}$ (up arrows) and spin down $\varepsilon_{+1/2}$ (down arrows) hybridized bands in URu$_{1.92}$Rh$_{0.08}$Si$_2$, with red line linear fits added to guide the eye. A pseudospin notation of $\pm 1/2$ is used for up- and down-spin states, respectively. The grey dot represents the approximation location of the quantum critical point or $H_{\rm m}$ obtained from published magnetotransport and magnetization measurements. (d) shows similar fitted values for CeRu$_2$Si$_2$.}
  \label{fig2}
\end{figure}

\begin{figure}[hctb]
  \centering
  \caption{Fitted values of $\Delta$ (the hybridized bandwidth) for both URu$_{1.92}$Rh$_{0.08}$Si$_2$ and CeRu$_2$Si$_2$, as indicated, plotted versus $H-H_{\rm m}$ with error bars. The dashed grey lines denote the regions of Fermi liquid and non-Fermi liquid recently identified from a crossover $T^*$ in the electrical resistivity~\cite{kim}. The vertical axes $\Delta$  and $T$ are scaled only by the Boltzmann constant $k_{\rm B}$, revealing a mutual consistency between experiments in which the Fermi temperature $T^*$ approximately corresponds to $\Delta$. The various coloured regions are labelled in accordance with Fig. 1, with the addition of a new phase (hashed region) which forms only in URu$_{1.92}$Rh$_{0.08}$Si$_2$ (not CeRu$_2$Si$_2$) as a means to avoid the putative quantum critical point.}
  \label{fig3}
\end{figure}

\begin{figure}[hctb]
  \centering
  \caption{Evidence for ordering in URu$_{1.92}$Rh$_{0.08}$Si$_2$. (a) compares actual $C_p/T$ data (diamonds) with a curve (solid line) calculated from parameters extrapolated to 37.5 T, obtained by fitting the resonance model for  $\mu_0H\geq$~38~T. The calculated curve matches the data for $T >$~6~K, but the difference (shaded regions) reveals a significant redistribution of entropy below 6 K. This firmly establishes the involvement of the 5$f$ doublets in ordering. (b) shows $C_p$ versus $T$ at several different magnetic fields (different colours) on both the initial $\approx$~0.1~K increase of the temperature (open symbols) and subsequent increases of temperature after cooling (filled symbols). The difference between them provides definitive evidence for hysteretic losses, which typically result from phase coexistence.}
  \label{fig4}
\end{figure}

\end{document}